# Status of Astronomy in Namibia

M. Backes[1,2,*], R. Evans[1], E.K. Kasai[1], and R. Steenkamp[1]

[1]*Department of Physics, University of Namibia, P/Bag 13301, Windhoek, Namibia*
[2]*Centre for Space Research, North-West University, P/Bag X6001, Potchefstroom, South Africa*
[*]corresponding author: mbackes@unam.na

**Abstract**

Astronomy plays a major role in the scientific landscape of Namibia and Southern Africa. Considerable progress has been achieved scientifically as well as in terms of human capacity development in the field. In all wavelength regimes accessible with ground-based instruments, the largest of those instruments are situated in Southern Africa: MeerKAT, the Southern African Large Telescope, and the High Energy Stereoscopic System. Because of the excellent observing conditions from Namibian soil, further large-scale projects such as the Cherenkov Telescope Array considered sites in Namibia and the Africa Millimetre Telescope will eventually be built there.

Against this background, the current situation of astronomical research and education in Namibia is reviewed, focusing on optical, radio and gamma-ray astronomy and also including smaller scale projects. Further, the role of astronomy, with particular focus on developmental aspects in the African context is outlined and the progress in human capacity development is summarized.

## 1. Introduction

In recent years, significant development in astronomy and space science has been accomplished in Africa. Southern Africa is becoming a beacon for astronomy: Throughout the electromagnetic spectrum the largest astronomical facilities are either operational or in the process of being set up in the region [1].

The Southern African Large Telescope (SALT) in Sutherland (South Africa), measuring 11 m in diameter, is the largest single optical telescope in the Southern hemisphere [2]. The deployment of the telescopes of the MeerKAT radio telescope, being the largest and most powerful radio telescope in the Southern hemisphere, has just been completed [3]. The 64-dish MeerKAT telescopes will later develop into the Square Kilometre Array (SKA), the most sensitive radio telescope on Earth, utilizing outlier station all over Southern Africa [4]. The High Energy Stereoscopic System (H.E.S.S.) telescopes [5] in the Khomas highlands in Namibia are the largest and most powerful system of Cherenkov telescopes to study very high energy ($E > 100$ GeV) gamma-rays. For its successor, the Cherenkov Telescope Array (CTA) [6,7], Namibia has been voted second of the possible countries to host the Southern part [8,9]. In the next sections the role of astronomy for development is be outlined and Namibia's comparative geographical advantage in terms of astronomy is discussed. We also summarize in the sections below the status of radio, optical, and gamma-ray astronomy in Namibia, as well as recent developments in Human Capacity Development (HCD).

## 2. The Role of Astronomy

Since ancient times astronomy is known as one of the major research areas pursued by mankind. And still today, it is a thriving force for innovative technologies and socioeconomic development, particularly in Africa [10]. Besides providing a skilled workforce that is used to handling Big Data and working in international collaborations, it has been shown that astronomy fosters general interest in science, technology, engineering and mathematics (STEM) subjects, at both primary and secondary school level. A recent study found that *"the one topic (among the sciences) that generated universal enthusiasm was any study of astronomy"* [11]. Hence, countries actively involved in astronomy not only earn the direct benefits from their involvement, but also attract more students into STEM fields, leading to higher innovativeness of their population.

The quest for ever deeper insights into the laws governing the Universe demand for ever larger astronomical facilities and for cutting-edge technologies, which eventually find their way into commercial products. The best-known examples are arguably charge-coupled devices (CCD-sensors), initially used for astronomy and nowadays found as image sensors in the cameras in most modern phones. Another example is WiFi, which uses algorithms developed by radio astronomers [12]. Figure 1 showcases the interconnection of astrophysics with different fields of engineering. From these strong interconnections, it becomes clear why astronomy is utilized by many countries as a drawing card for sustainable development of the whole Science and Technology sector, as pointed out e.g. by the (then) Minister of Science and Technology of South Africa [13].

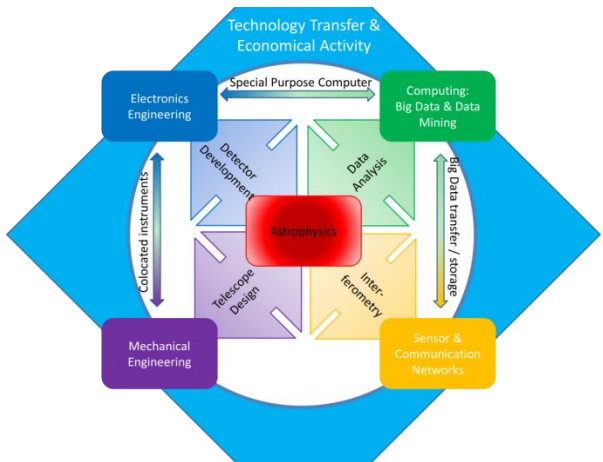

**Figure 1:** Interconnections of astrophysics with different fields of engineering.

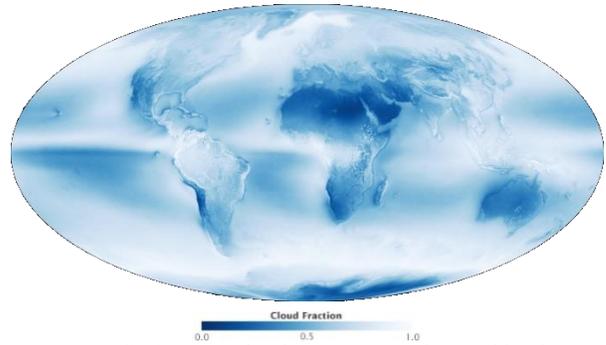

**Figure 3:** Average cloud coverage as observed by the MODIS satellite between July 2002 and April 2015. Colors range from dark blue (no clouds) to light blue (some clouds) to white (frequent clouds) [16].

## 3. Namibia - Geographically Advantaged for Astronomy

Being situated in the Southern hemisphere, around the Tropic of Capricorn at -23.4° latitude, Namibia enjoys a direct overhead view of the central region of the Milky Way during the astronomically preferred winter nights. Whereas this is common to all countries around the same latitude, Namibia, in addition, offers some of the darkest, least light polluted skies around the globe, as illustrated in Figure 2 and discussed in [15].

Furthermore, major parts of Namibia are elevated 1,500 to 2,500 m a.s.l. and constitute some of least cloudy places in the Southern hemisphere, as illustrated in Figure 3. Detailed investigation of nightly cloud coverage in Southern Africa, conducted in the context of CTA, support this global picture [17].

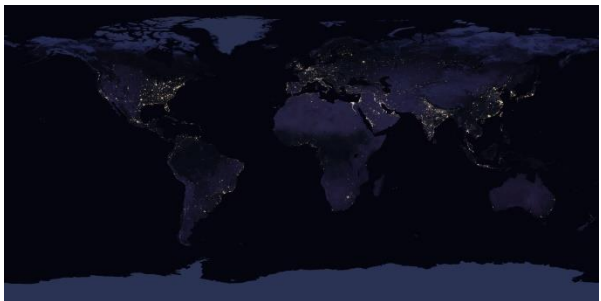

**Figure 2:** Night-lights recorded by NASA's Earth Observatory in 2016 [14].

For this and the world-leading astronomical seeing conditions on Mt. Gamsberg [18], there has been a long-standing plan for a large-scale astronomical observatory in Namibia, see e.g. [19]. Meanwhile, the H.E.S.S. telescopes have been built and the SKA will build several outlier stations in Namibia. Hence, there are very positive prospects for multi-wavelength astronomy in Namibia. In the following, the current situation in radio, optical, and gamma-ray astronomy in Namibia will be summarized.

## 4. Radio Astronomy

Radio astronomy in Southern Africa is mostly focused towards the SKA and its pathfinder MeerKAT [20]. In August 2017, the Namibian Minister of Higher Education, Training and Innovation signed the Memorandum of Understanding for Institutionalizing Cooperation in Radio Astronomy, together with representatives of the other 8 African SKA Africa partner countries [21]. In the course of that, a strong emphasis is made to human capacity development through projects like the African VLBI Network (AVN) [22], which is also financially supported by the UK-SA Newton Fund [23].

Training the first generation of radio astronomers in the SKA partner countries is the aim of the Development in Africa with Radio Astronomy (DARA) project [24]. Since 2017, yearly training sessions of 5 Namibian graduates are organized collaboratively by the Universities of Namibia (UNAM), Leeds, and Oxford, the Botswana International University of Science and Technology, and the South African Hartebeesthoek Radio Astronomy Observatory (HartRAO).

Also, efforts are being undertaken to set up a high-performance computing cluster in Namibia, supported by the Centre for High Performance Computing (CHPC) in Cape Town, and making use of part of the former RANGER cluster sponsored by the Texas Advanced Computing Center (TACC) [25].

Recently, it was proposed to complement the Event Horizon Telescope (EHT), a world-wide very long baseline interferometry (VLBI) array of millimeter-wave radio telescopes with one telescope in Namibia [26]. The main scientific goal of the EHT is to resolve the shadow of the black hole at the center of the Milky Way Galaxy, Sagittarius A*. By adding a telescope in Southern Africa to this VLBI network, significant improvements in image quality and, hence, angular resolution are expected [27]. Mt. Gamsberg in Namibia has been chosen as the ideal location for this first of its kind telescope in Africa, dubbed the Africa Millimetre Telescope.

## 5. Optical Astronomy

As previously mentioned, Mt. Gamsberg has attracted significant international attention since the 1970s for building an international observatory. Early attempts [18] were followed by considering it as a possible site for SALT [28]. To date, the International Amateur Observatory (IAS) [29] operates a 71 cm telescope on Mt. Gamsberg [30] which currently is the second largest optical telescope in Namibia. For a recent review of the astronomical observing conditions on Mt. Gamsberg see [26].

The optical telescope with the largest aperture currently operating in Namibia is the ATOM telescope [31] on the H.E.S.S. site. Its main purpose is the optical monitoring support of H.E.S.S. observations but also occultation observations are conducted with it. Furthermore, from the H.E.S.S. site, the HATSouth Exoplanet Survey is performed [32]. Besides these (semi-) professional projects, Namibia hosts a vivid community of amateur astronomers, fostering the sector of astro-tourism. One of these private observatories, the Cuno Hoffmeister Memorial Observatory (CHMO), is located 15 km South of Windhoek and hence well suited for educational purposes. Its main instrument is a 14" (f/11) Schmidt-Cassegrain telescope which was recently equipped with a SBIG STF-8300M CCD camera, supported by the International Astronomical Union (IAU) Office of Astronomy for Development (OAD) [33], for the purpose of performing photometric observations, e.g. of globular clusters. Additionally, this observatory is used for occultation astronomy, observing trans-Neptunian objects (TNOs), in collaboration with the International Occultation Timing Association European Section (IOTA-ES) and the Observatoire de Paris [34,35].

The Department of Physics at UNAM has been involved using SALT in follow-up spectroscopic observations of supernova candidates discovered by the international Dark Energy Survey Supernova (DES-SN) programme, since the second-half of 2016 until earlier this year, when the DES-SN programme came to an end. However, the SALT follow-up spectroscopy of DES supernova candidates began in 2013 when a member of staff at UNAM involved in the work was at that time a PhD student at the University of Cape Town, the South African Astronomical Observatory and the African Institute for Mathematical Sciences [36].

## 6. Gamma-ray Astronomy

The H.E.S.S. telescopes have been successfully operating in Namibia since 2002, with the inauguration of a fifth 28-m diameter telescope, dubbed H.E.S.S.-II in 2012 [37,38]. Namibian involvement in H.E.S.S. has geared up significantly in the last few years, with several students having served as telescope operators. Three undergraduate and one postgraduate students have analyzed some of the data gathered by the H.E.S.S. telescopes and one postgraduate student was involved in a major technical upgrade of the cameras [39,40]. Furthermore, two UNAM staff members are currently pursuing their PhDs in the field of multi-wavelength and gamma-ray astronomy with the H.E.S.S. telescopes, e.g. [41,42]. Another postgraduate student is studying the front-end electronics for the Compact High Energy Camera (CHEC) for the small-size telescopes of CTA [43]. Namibia's strong initiative in bidding to host the CTA [8] was put on hold in 2015, as the CTA Resource Board (the governing body of CTA) decided to first start negotiations with ESO for a site in Chile [9]. The scientists involved in the CTA had previously found the Namibian site to have the highest scientific merit world-wide [44].

## 7. Human Capacity Development

Recent HCD initiatives like the 1st Namibian JEDI Workshop on Astronomy and Space Science [45], the DARA project as explained above, and the recent setting-up of a HiSPARC [46] cosmic ray detector at UNAM have already shown positive effects on the number of project students in astrophysics, as shown per subfield in Table 1 and per year in Figure 4. These numbers are expected to rise further once the Namibia/South Africa bilateral research chair in Astronomy becomes operational later this year.

**Table 1**: Number of current and recent (since 2013) staff and students in the different fields of astronomy at UNAM.

| Field | BSc (Hons) stud. | MSc stud. | PhD stud. | Staff with PhD /MSc |
|---|---|---|---|---|
| Radio Astronomy | 4 | 1 | - | 1 / 0 |
| Optical Astronomy | 10 | 1 | 1 (SA) | 1 / 0 |
| Gamma-ray Astronomy | 4 | 4 | 2 (SA) | 1 / 3 |
| Theoretical Astronomy | 4 | 0 | 0 | 1 / 0 |

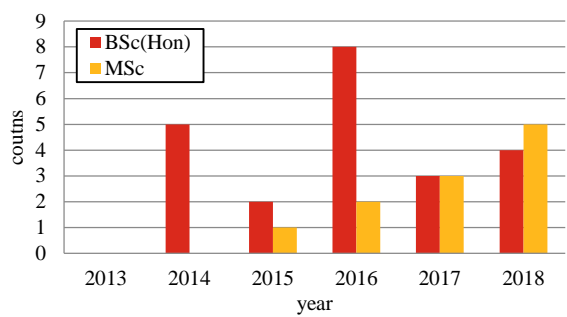

**Figure 4:** Student enrollment at UNAM in project year of BSc(Hons) in Physics and MSc in Physics, both with astrophysics projects.

## 8. Conclusions

Assessing the status of astronomy in Namibia as presented above, including the tremendous progress the Namibian government has made towards astronomy development in the country, one can safely conclude that a very good starting point has been established. It is particularly worth mentioning that "Space Science", including basic astronomical research, has been singled out as one of the *Key Research Areas* by the National Commission on Research, Science & Technology [47] and that a National (advisory) Council for Space Science has been set up.

## Acknowledgements

Partial support by the National Commission on Research, Science, and Technology, the International Astronomical Union Office of Astronomy for Development, the UK-SA Newton Fund, the Global Challenges Research Fund, the Africa Oxford Initiative, the Institute of Physics, and the International Centre for Theoretical Physics is gratefully acknowledged.